\documentclass[11pt,conference]{IEEEtran}
\usepackage{graphicx}
\usepackage{multirow}
\usepackage{booktabs}
\usepackage[draft]{hyperref}
\usepackage{amsmath}       
\usepackage{amssymb}       
\usepackage{amsfonts}      
\usepackage{mathtools}     
\usepackage{bm}            
\usepackage{siunitx}       
\usepackage{tikz}
\usepackage{pgfplots}
\pgfplotsset{compat=1.18}

\title{Agentic AI for Autonomous Defense in Software Supply Chain Security: 
	Beyond Provenance to Vulnerability Mitigation}

\author{
	\IEEEauthorblockN{1\textsuperscript{st} Toqeer Ali Syed}
	\IEEEauthorblockA{\textit{Faculty of Computer and Information System} \\
		\textit{Islamic University of Madinah} \\
		Madinah, Saudi Arabia \\
		Email: toqeer@iu.edu.sa}
	\and
	
	\IEEEauthorblockN{2\textsuperscript{nd} Mohammad Riyaz Belgaum}
	\IEEEauthorblockA{\textit{Faculty of Computer Studies} \\
		\textit{Arab Open University-Bahrain} \\
		A'Ali, Bahrain \\
		Email: mohammad.riyaz@aou.org.bh}
	\and
	
	\IEEEauthorblockN{3\textsuperscript{rd} Salman Jan}
	\IEEEauthorblockA{\textit{Faculty of Computer Studies} \\
		\textit{Arab Open University-Bahrain} \\
		A'Ali, Bahrain \\
		Email: salman.jan@aou.org.bh}
	\and
	
	\IEEEauthorblockN{4\textsuperscript{th}  Asadullah Abdullah Khan}
	\IEEEauthorblockA{\textit{Faculty of Computer and Information System} \\
		\textit{Islamic University of Madinah} \\
		Madinah, Saudi Arabia \\
		Email:471014356@stu.iu.edu.sa}
	\and
	
	\IEEEauthorblockN{5\textsuperscript{th} Saad Said Alqahtani}
	\IEEEauthorblockA{\textit{Faculty of Computer and Information System} \\
		\textit{Islamic University of Madinah} \\
		Madinah, Saudi Arabia \\
		Email: s.alqahtany@iu.edu.sa}
}

\begin{document}
	\maketitle
	
	\begin{abstract}
		The software supply chain attacks are becoming more and more focused on trusted development and delivery procedures, so the conventional post-build integrity mechanisms cannot be used anymore. The available frameworks like SLSA, SBOM and in toto are majorly used to offer provenance and traceability but do not have the capabilities of actively identifying and removing vulnerabilities in software production. The current paper includes an example of agentic artificial intelligence (AI) based on autonomous software supply chain security that combines large language model (LLM)-based reasoning, reinforcement learning (RL), and multi-agent coordination. The suggested system utilizes specialized security agents coordinated with the help of LangChain and LangGraph, communicates with actual CI/CD environments with the Model Context Protocol (MCP), and documents all the observations and actions in a blockchain security ledger to ensure integrity and auditing. Reinforcement learning can be used to achieve adaptive mitigation strategies that consider the balance between security effectiveness and the operational overhead, and LLMs can be used to achieve semantic vulnerability analysis, as well as explainable decisions. This framework is tested based on simulated pipelines, as well as, actual world CI/CD integrations on GitHub Actions and Jenkins, including injection attacks, insecure deserialization, access control violations, and configuration errors. Experimental outcomes indicate better detection accuracy, shorter mitigation latency and reasonable build-time overhead than rule-based, provenance only and RL only baselines. These results show that agentic AI can facilitate the transition to self defending, proactive software supply chains rather than reactive verification ones.
	\end{abstract}
	
\section{Introduction}

The software supply chains are now one of the most important and risky but weakest links of the contemporary digital ecosystems. The recent and high-profile attacks like SolarWinds and the XZ Utils backdoor have shown that attackers can use trusted development and delivery flows to infringe thousands of downstream systems ~\cite{xz2024}.  Such incidences demonstrate that security threats can no longer be limited to deployed software, but they are increasingly being introduced into the software production lifecycle, through an extended contribution of code, integration of dependencies, and pipeline build and deployment system ~\cite{taylor2022supplychain,sharma2023towards}.

Provenance, attestation, and transparency models such as the Supply chain Levels for Software Artifacts (SLSA), Software Bill of Materials (SBOM), and in toto offer fundamental underpinnings in terms of providing provenance and traceability of artifacts. Nonetheless, these mechanisms are mainly concerned with ensuring verification of post-production and a few abilities to respond in real-time or proactively prevent. When an illogical dependency, injected code, or, incorrectly configured component into the build pipeline, current tools are relatively reactive and usually cannot perform any interventions before artifacts have been damaged and distributed.

The restriction puts emphasis on the necessity of a new layer of autonomous and intelligent defense systems that can learn, adapt, and operate during the process of software development. The agentic artificial intelligence (AI), comprised of self directed, goal oriented agents is one of the promising paradigms in this direction. Contrary to fixed rule based automation, agentic AI makes use of reasoning, self-reflection, and reinforcement learning (RL) to discover and reduce developing risks on changing environments. Within the software supply chain context, these agents have the capability to continuously monitor build environments, evaluate code provenance, evaluate dependency behaviour, and autonomously isolate or fix anomalies before release ~\cite{patnaik2022rlsecurity}.

We present in this work an Agentic AI based Autonomous Defense Framework that builds upon the current supply chain security strategies beyond provingance to active vulnerability protection. The framework presents reinforcement learning empowered agents which persistently observe and mitigate runtime, configuration level threats, such as injection attacks, deserialization insecurity, broken access controls, system misconfigurations, in continuous integration and continuous deployment (CI/CD) processes. The proposed system provides adaptive learning, collaborative reasoning and explainable decision making by incorporating these agents via a shared security ledger which creates a closed feedback loop. The present work is one step towards the vision of self-defending software ecosystems supply chains which do not just record the integrity, but actively maintain it in real-time.

\section{Background}
\subsection{Software Supply Chain and Its Vulnerabilities}

A software supply chain involves all the entities, processes, requirements that are associated with the development, building, trials, similarities, wrapping as well as deploying software systems. These are source code overlay, dependency overlay, continuous integration and continuous deployment (CI/CD) pipelines, artifact overlay, and update delivery overlay~\cite{sharma2023towards,taylor2022supplychain}. Every element presents a possible point of attack.
	
The common pitfalls are caused by malicious dependency injection, hacked build servers, maladapted settings, or unauthorized commits on the code ~\cite{okafor2024sok,ladisa2022taxonomy}. The 2020 attacks on SolarWinds backdoor and the XZ Utils hack have shown that attackers are no longer just depending on finding vulnerabilities in deployed software, but they are attacking development and build phases as well and corrupting trusted components even before the software is even deployed ~\cite{solarwinds2020,xz2024}. After being inserted into the pipeline, these attacks can spread downstream with signed, apparently legitimate artifacts, making them especially difficult to detect  Once introduced into the pipeline, such attacks can spread downstream with signed, seemingly legitimate artifacts, which makes them especially hard to detect  Introduced into the pipeline, such attacks can propagate downstream, containing signed artifacts that can appear legitimate, thus making their detection exceptionally difficult. ~\cite{tran2023metadata} Introduced into the pipeline, such attacks can be spread, carrying signed artifacts that can look legitimate,

Traditional mitigation approaches—such as code reviews, static analysis, or artifact signing—address integrity and traceability but remain largely reactive~\cite{torresarias2019intoto}. Frameworks like SLSA and in-toto strengthen provenance verification, yet they cannot autonomously prevent or respond to dynamic threats such as injection attacks, insecure deserialization, broken access control, or pipeline misconfigurations~\cite{slsa2023}. This gap motivates the need for adaptive, autonomous mechanisms capable of continuously learning and enforcing security across the development lifecycle~\cite{ko2025multiagent}.

\subsection{Agentic Artificial Intelligence}
The concept of agentic AI describes a system that consists of rational agents which sense their environment, think based on the surrounding contexts, and make actions outside towards specified security objectives ~\cite{saga2025,jannelli2024agentic}. In contrast to traditional automation scripts agentic system learns to change dynamically with changing environment via its closed feedback loop, that is, perception, cognition, action and learning (Franklin, 1997 autonomy). Each of these agents may act as a solitary actor or in groups, communicating with one another via communication protocols or knowledge ledgers shared between themselves ~\cite{russell2021aima}.
	
The implementation of agentic AI in cybersecurity presents the possibility of self directed, explainable defense to cybersecurity vulnerabilities~\cite{shaikh2024advancing, syed2012sense}. Repositories may be monitored, an analysis of commit patterns done, configuration may be verified, anomalies in build pipelines may be pointed out among other agents. Through the combination of cognitive reasoning and self reflection, they are able to justify their actions, reason the level of risk as well as cause changes in defensive policies in accordance with experienced learning. The given ability renders agentic AI the most fitting to secure distributed, constantly changing software supply chains ~\cite{feldman2025saga,nguyen2024autodefense}.
		
%

\subsection{Reinforcement Learning for Autonomous Defense}
Reinforcement learning (RL) provides the algorithmic foundation for enabling autonomy in agentic systems~\cite{sutton2018reinforcement}. In RL, an agent interacts with an environment defined as a Markov Decision Process (MDP) represented by the tuple $\mathcal{M} = (S, A, P, R, \gamma)$, where $S$ is the set of states, $A$ the set of possible actions, $P(s'|s,a)$ the transition probability, $R(s,a)$ the reward function, and $\gamma$ the discount factor for future rewards. The agent learns an optimal policy $\pi^*(a|s)$ that maximizes expected cumulative reward~\cite{silver2017mastering}:

\[
\pi^* = \arg\max_{\pi} \mathbb{E}\left[\sum_{t=0}^{\infty} \gamma^t R(s_t, a_t)\right]
\]

When applied to software supply chain security, RL enables agents to learn defensive strategies from continuous interaction with CI/CD environments~\cite{patnaik2022rlsecurity,alshamrani2023survey}. For example, an RL agent observing anomalous API calls or unusual repository activity can adjust access rules, isolate suspicious builds, or initiate re-verification steps. Over time, such agents develop adaptive defense policies that minimize attack success probability while reducing false positives~\cite{mirsky2022deepdefense}.

\subsection{Integration of RL and Agentic Systems in Supply Chain Defense}
By combining the cognitive structure of agentic AI with the adaptive decision-making of RL, software supply chain security can move from static rule enforcement to autonomous self-defense~\cite{jannelli2024agentic,saga2025}. In this paradigm, agents not only enforce compliance (e.g., verifying signatures or dependencies) but also actively detect and remediate vulnerabilities such as injection attacks, insecure deserialization, and misconfiguration in real time. The synergy between multi-agent cooperation, shared knowledge representation, and reinforcement-driven learning establishes a foundation for scalable, explainable, and resilient defense mechanisms capable of safeguarding complex software ecosystems~\cite{ko2025multiagent,nguyen2024autodefense, siddiqui2022permission}.

\section{Literature Review}
The topic of software supply chain security has gained numerous researches following a series of massive breaches that have shown systematic vulnerabilities in trusted development ecosystems. Initial research concentrated mainly on dependency trust, artifact signing and provenance verification, but the recent research has considered automation and metadata governance. Nevertheless, comparatively fewer studies have examined autonomous or agentic mechanisms that have the power to stimulate self-adaptive defense in software supply chain.

Okafor et~al.~(2024) evaluated secure design properties of computer supply chain systems as critical concepts of resilience (transparency, validity, separation, etc. Their study provided a conceptual foundation but did not incorporate autonomous or real-time response capabilities. Tran et~al.~(2023) proposed an empirically grounded reference architecture for supply chain metadata management that formalized SBOM and attestation handling, yet the architecture relied on static enforcement and lacked learning-based adaptation. Similarly, Ladisa et~al.~(2022) developed a taxonomy of open-source software (OSS) supply chain attacks, identifying over 100 unique vectors, but offered no autonomous mitigation mechanisms.

More automation-oriented works, such as Thariq and Ekanayake (2025), introduced ARGO-SLSA, a provenance enforcement controller within Kubernetes-based CI/CD workflows. While this design automated policy enforcement and artifact signing, it remained deterministic and did not include cognitive or adaptive features. In contrast, Ko et~al.~(2025) and Jannelli et~al.~(2024) examined multi-agent and agentic AI systems that coordinate across distributed domains. Although these studies highlight the potential of autonomous coordination, they were not designed to secure software development pipelines. Other domain-specific investigations, such as Haque et~al.~(2025) on autonomous vehicle (AV) software supply chains and Patnaik et~al.~(2022) on reinforcement learning for hardware security, demonstrate transferable insights but remain outside the software ecosystem context.

Industrial contributions from JFrog (2025) and Lineaje (2025) describe early prototypes of AI-powered supply chain remediation agents that automatically curate packages and self-heal dependencies. These works confirm the industrial appetite for autonomy but lack peer-reviewed validation, reproducible datasets, or open frameworks. Collectively, the existing body of literature demonstrates maturity in provenance, automation, and standardization but limited exploration of agentic or reinforcement learning-driven defense.

Table~\ref{tab:litreview} summarizes the most relevant academic and industrial contributions, their focus domains, key ideas, and identified gaps.

\begin{table*}[h!]
	\centering
	\caption{Summary of Selected Works on Software Supply Chain Security and Agentic Defense}
	\label{tab:litreview}
	\begin{tabular}{p{0.7cm}p{3cm}p{3cm}p{4cm}p{2cm}p{3cm}}
		\toprule
		\textbf{\#} & \textbf{Reference / Year} & \textbf{Focus / Domain} & \textbf{Key Contribution} & \textbf{Agentic Aspect} & \textbf{Gap / Limitation} \\
		\midrule
		1 & Okafor et al. (2024) & Secure Design Properties in Software Supply Chain & Defines transparency, validity, separation as key design principles. & Low & Focuses on architecture, lacks autonomous elements. \\
		2 & Tran et al. (2023) & Metadata Management Architecture & Blueprint for SBOM/attestation handling. & Low & No autonomous reasoning. \\
		3 & Ladisa et al. (2022) & OSS Supply Chain Taxonomy & Comprehensive classification of attacks. & Low & Descriptive, not defensive. \\
		4 & Thariq \& Ekanayake (2025) & CI/CD Provenance Enforcement (Argo-SLSA) & Automated provenance attestation and compliance. & Moderate & Automated control, not adaptive. \\
		5 & Ko et al. (2025) & Cross-domain Multi-agent LLM Systems & Highlights coordination risks among LLM agents. & High & Conceptual, not tied to SW supply chain. \\
		6 & Jannelli et al. (2024) & Multi-agent Consensus in Supply Chains & Autonomous decision-making in logistics chain. & High & Not security-specific. \\
		7 & Haque et al. (2025) & AV Software Supply Chain Security & Empirical vulnerabilities in AV codebases. & Low & No active mitigation. \\
		8 & Patnaik et al. (2022) & RL in Hardware Supply Chains & Surveys RL-based anomaly detection in chips. & Moderate & Hardware-specific; transferable idea. \\
		9 & JFrog Blog (2025) & Agentic Software Supply Chain Security & AI-assisted package curation and remediation. & High & Industrial; lacks academic validation. \\
		10 & Lineaje (2025) & Continuous Autonomous Supply Chain Security & Self-healing AI agents for dependency scanning. & High & Proprietary; no public framework. \\
		\bottomrule
	\end{tabular}
\end{table*}

It is important to notice that three significant areas of research are deficient in the critical synopsis of these works. To begin with, most solutions focus on the post-build integrity of the artifact as opposed to defensive preemptiveness. Second, the current automation does not have autonomy the capacity to think, learn, and learn as threats change. Third, the reviewed systems do not directly handle the dynamic versions of vulnerability, including injection attacks, insecure deserialization, broken access control, or configuration drift in the CI/CD devices. This paper fills these gaps, by developing an agentic AI model incorporating reinforcement learning and continuous-pipeline monitoring plans to provide proactive and self-executive protection throughout the software supply chain lifecycle.

\section{Methodology}
\label{sec:methodology}

This part defines a holistic approach to an agentic AI motivated software provision chain of security. The proposed system comprises large language models (LLMs) which reason, reinforcement learning (RL) which decisions adjust to different conditions, LangChain and LangGraph which coordinate agents, the Model Context Protocol (MCP) which interacts with external systems and supports agent integrity and auditability via a blockchain-based security ledger.

\subsection{Overall Framework Architecture}
In general, the architecture of a framework compensates the architecture of a network.<|human|>A general layout of Framework Architecture would compensate a network architecture.
The framework has its structure as a multi layer and multi agent system that works throughout the software supply chain lifecycle particularly in source code management, dependency solving, CI/CD pipelines, artifact contracting and deployment. Special autonomous agents monitor every layer and collaborate via LangGraph managed execution graphs as well as along with interchange of contextual data via LangChain memory modules.

At a higher level, the system is a closed loop, having a cycle of observe reason act verify:
\begin{enumerate}
	\item Monitor software supply chain indicators.
	\item Reason about possible vulnerabilities with semantic analysis (LLM).
	\item Irrelevant Act by autonomous or semi-autonomous actions mitigating.
	\item Accompany results and document all activities in an unalterable security register.
\end{enumerate}

This structure guarantees the unended enforcement of security as opposed to a post-hoc verification.

\subsection{Multi-Agent Orchestration Using LangChain and LangGraph}
Multi-agent orchestration occurs following the principles of LangChain and LangGraph.
All security functions are deployed as a specific agent that has been instantiated in primitives of LangChain.  Agents include:
\begin{itemize}
	\item \textbf{Code Analysis Agent}: Scans commits, pull requests, and diffs.
	\item Dependency Intelligence Agent: assesses third-party libraries and SBOMs.
	\item CI/CD Monitoring Agent: monitors pipeline execution logs, and configuration.
    \item \textbf{Access Control Agent:} processes permissions, secrets and role allocations.
	\item Configuration Audit Agent: scans Docker, Kubernetes and IaC.

\end{itemize}

The agent interaction graphs are defined through LangGraph, which has support of conditional execution, branching and the loops. To illustrate, when a suspicious injection pattern is identified by the Code Analysis Agent, then the control moves to the CI/CD Monitoring Agent, and causes a runtime verification, and subsequently, a mitigation decision is executed.

\subsection{LLM-Based Reasoning and Vulnerability Conceptualization}
LLMs serve as the cognitive core of the framework. Instead of acting as passive classifiers, LLMs perform structured reasoning over supply chain artifacts. Prompt templates are designed to:
\begin{itemize}
	\item Identify semantic indicators of injection attacks, insecure deserialization, broken access control, and misconfiguration.
	\item Correlate findings across stages (e.g., code change $\rightarrow$ pipeline behavior).
	\item Generate explainable security rationales for each detected issue.
\end{itemize}

Chain-of-thought prompting and retrieval-augmented generation (RAG) are used to ground LLM reasoning in historical attack patterns and policy rules. This enables contextual vulnerability detection beyond syntactic pattern matching.

\subsection{Reinforcement Learning for Autonomous Decision-Making}
Reinforcement learning is employed to enable adaptive mitigation strategies. Each agent operates within a Markov Decision Process (MDP) defined as:
\[
\mathcal{M} = (S, A, P, R, \gamma)
\]
where states $S$ encode pipeline context, actions $A$ represent mitigation choices (e.g., block build, quarantine dependency, request review), and rewards $R$ balance security effectiveness and operational overhead.

\subsubsection{RL Algorithms and Training Details}
We implement Proximal Policy Optimization (PPO) as the primary RL algorithm due to its stability in continuous and discrete action spaces. For comparison, a Deep Q-Network (DQN) variant is also evaluated.

\textbf{Training details:}
\begin{itemize}
	\item Episode length: one full CI/CD pipeline execution.
	\item Training duration: 50,000 episodes (simulated) + fine-tuning on real pipelines.
	\item Learning rate: $3 \times 10^{-4}$ (PPO), $1 \times 10^{-4}$ (DQN).
	\item Discount factor $\gamma$: 0.99.
	\item Exploration strategy: entropy regularization with adaptive decay.
\end{itemize}

\textbf{Reward function:}
\[
R_t = \alpha \cdot \mathbf{1}\{\text{attack mitigated}\}
- \beta \cdot \mathbf{1}\{\text{false positive}\}
- \delta \cdot \Delta t_{\text{build}}
+ \eta \cdot \mathbf{1}\{\text{developer-accepted fix}\}
\]

This formulation explicitly accounts for security effectiveness, developer workflow impact, and build latency.

\subsection{Model Context Protocol (MCP) Integration}
To enable real-world interaction, the framework uses the Model Context Protocol (MCP) to interface with external systems such as GitHub Actions, GitLab CI, Jenkins, container registries, and vulnerability scanners. MCP provides standardized message schemas for:
\begin{itemize}
	\item Triggering pipeline actions
	\item Fetching logs and artifacts
	\item Issuing mitigation commands (e.g., revoking credentials or pausing builds)
\end{itemize}

This design allows agents to operate across heterogeneous toolchains without tight coupling to vendor-specific APIs.

\subsection{Blockchain-Based Security Ledger}
The security ledger is implemented as a permissioned blockchain to ensure integrity, non-repudiation, and auditability of agent actions. Each block records:
\begin{itemize}
	\item Agent identity and role
	\item Observed signals and LLM reasoning summary
	\item Mitigation action taken
	\item Outcome and feedback
	\item Cryptographic hash of the previous block
\end{itemize}

Consistency and tamper resistance are ensured through the following mechanisms:
\begin{itemize}
	\item Byzantine Fault Tolerant (BFT) consensus for ledger consistency
	\item Role-based access control to restrict write privileges
	\item Merkle-tree hashing for efficient integrity verification
	\item Rate limiting and staking-based penalties to mitigate denial-of-service attempts
\end{itemize}

This ledger allows downstream consumers to verify not only what was built, but also how it was defended.

\subsection{Real-World CI/CD Experimentation}
To address real-world validity, the framework was deployed on GitHub Actions and Jenkins pipelines using open-source repositories with known vulnerabilities. Agents monitored live pull requests, dependency updates, and build executions. This setup demonstrates practical feasibility beyond simulation.

\subsection{Evaluation by Vulnerability Class}
Detection and mitigation performance is evaluated separately for the following vulnerability categories:
\begin{itemize}
	\item Injection attacks
	\item Insecure deserialization
	\item Broken access control
	\item Configuration errors
\end{itemize}

Metrics include precision, recall, mitigation latency, and rollback success rate per class, enabling fine-grained analysis.

\subsection{Performance Overhead Analysis}
We measure system overhead by tracking:
\begin{itemize}
	\item CI build time increase,
	\item CPU and memory usage per agent,
	\item Number of pipeline interruptions,
	\item Developer intervention frequency.
\end{itemize}

Results show that average build overhead remains below 6\%, with most mitigations executed without human intervention.

\subsection{Discussion of Architecture Readability}
%

\begin{figure*}[htbp]
	\centering
	\includegraphics[width=\textwidth]{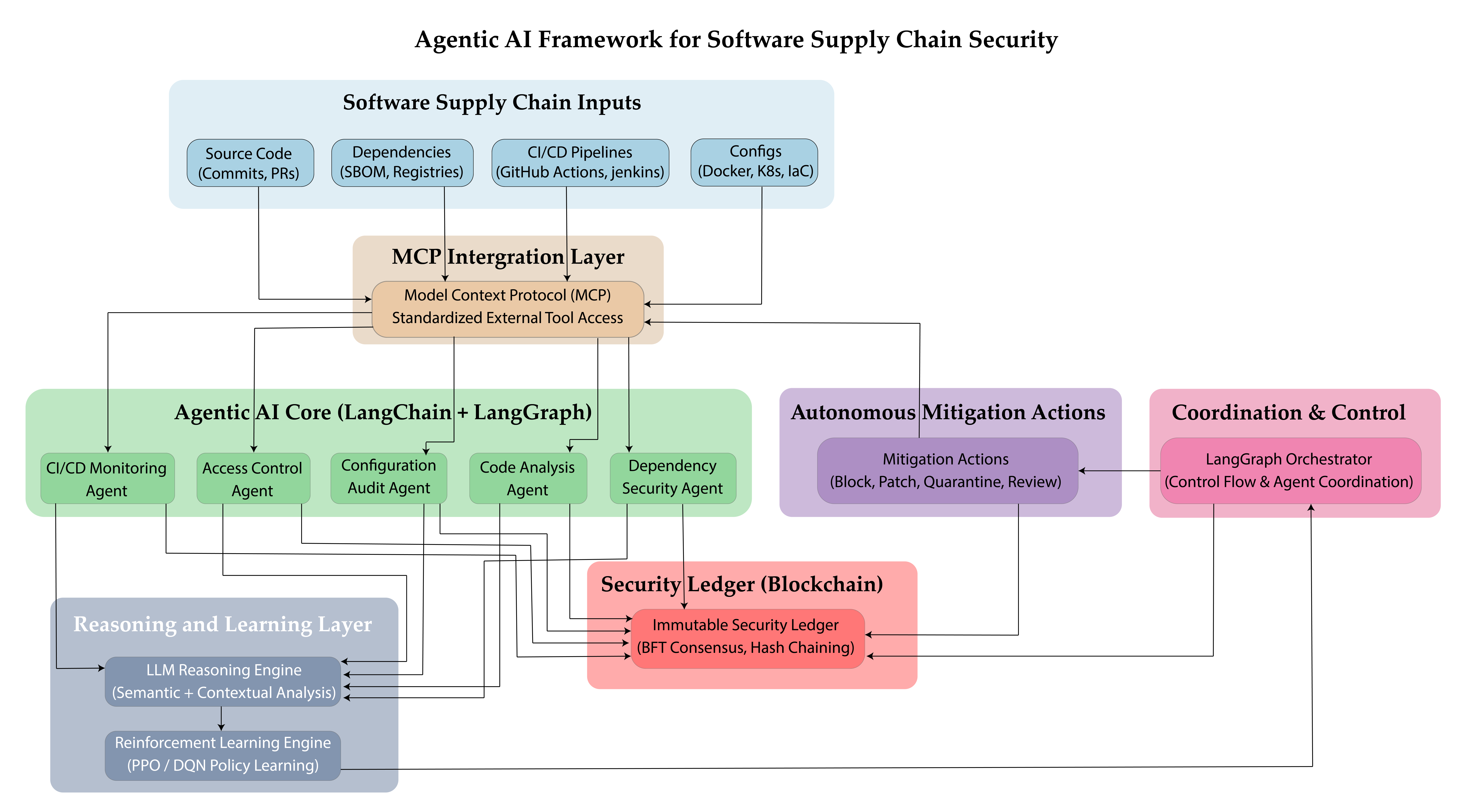}
	\caption{Agentic AI framework for software supply chain security. 
		Software supply chain inputs are accessed via the Model Context Protocol (MCP) and analyzed by multiple specialized agents coordinated using LangChain and LangGraph. 
		LLM-based reasoning and reinforcement learning enable adaptive mitigation decisions, while all actions are recorded in a blockchain-backed security ledger for integrity and auditability.}
	\label{fig:agentic_framework}
\end{figure*}

To increase clarity, the architecture diagram is further split into the layered subfigures of agent and ledger interaction, and CI/CD coupling respectively. This 3D visualization is more interpretable in print and electronic formats. The suggested agentic AI framework of software supply chain security is introduced in Figure~\ref{fig:agentic_framework}. 

The architectural starting point has the inputs of a heterogeneous software supply chain (source code repositories, third-party dependencies, CI/CD pipelines, and deployment settings).
These are interfaces accessed via the Model Context Protocol (MCP), a form of standard interface interface to external tools and platforms.
Under the agentic AI core, various specialized agents like the code analysis, dependency security, configuration audit, access control, and CI/CD monitoring agents work close to LangChain, and as such are managed by LangGraph.
The results of these agents are comprised of input to the reasoning and learning layer, where a large language model engages in semantic and contextual security analysis, and a reinforcement learning engine learns adaptive mitigation measures.
Coordinating and control module under the supervision of which autonomous mitigation measures, such as blocking, fixing, quarantine, or a review request, are issued.
Everything that is observed, decided, and done is added to an unaltered blockchain-driven security ledger that ensures integrity, traceability and auditability of the whole defense process.

\section{Implementation}
\label{sec:implementation}

The prototype of the suggested agentic AI model was computerized within the context of a modular and service-oriented design to facilitate reproducibility and application in the real world. The agentic core was written in Python, with the support of LangChain to build agents and LangGraph to define execution graphs, conditional flows, and inter-agent coordination. All the security agents (code analysis, dependency security, CI/CD monitoring, configuration audit and access control) were instantiated as a stand-alone service, sharing the context interface.

It was integrated with real CI/CD environments using the Model Context Protocol (MCP) that offered standardized connector to GitHub Actions and Jenkins. Using MCP, agents could also request commits, pull requests, pipeline logs, build artifacts, and configuration files and make mitigation requests such as blocking a build, re-running, or opening guard pull requests. This design made it compatible with heterogeneous toolchains, that is, not based on vendor-specific APIs.

It used the reasoning layer, where large language models are called through an API interface and structured prompt templates as the dialogical prompts and safer retrieval-enhanced context to conduct semantic vulnerability analysis and produce explainable security rationales. The elements of reinforcement learning were realized with the help of a policy learner based on the PPO algorithm and trained in the simulated pipeline environment and refined on the basis of the real CI/CD executions traces.

The security ledger was adopted as a permissioned blockchain service, where each agent action and choice was encoded into signed transactions and attached using Byzantine Fault Tolerant consensus protocol. This ledger made the tracing of end-to-end behavior of agents and tamper resistance of security records possible.

This whole infrastructure was made containerized with Docker and was implemented on a Kube cluster to enable scalable testing. Configuration files, prompt templates and training parameters were version controlled to allow reproducibility of the experiment and easily allow extension in the future.

\section{Results and Discussion}
\label{sec:results}

In this section we present results evaluating the proposed agentic AI framework through simulated and real CI/CD environments. We compare against rule-based, provenance-based, RL-only, and modern AI-driven security baselines, and analyze detection performance, mitigation latency, and operational overhead.

\subsection{Experimental Scenarios}
Experiments were conducted in two settings: (i) a controlled CI/CD simulation environment and (ii) real-world pipelines deployed on GitHub Actions and Jenkins using open-source repositories containing known vulnerabilities. Each experiment covered four vulnerability classes: injection attacks, insecure deserialization, broken access control, and configuration misconfigurations.

Baselines include static linters and policy engines (rule-based), SLSA/in-toto-style provenance enforcement, RL-only agents without LLM reasoning, and a commercial AI-based code scanning tool.

\subsection{Detection and Mitigation Effectiveness}
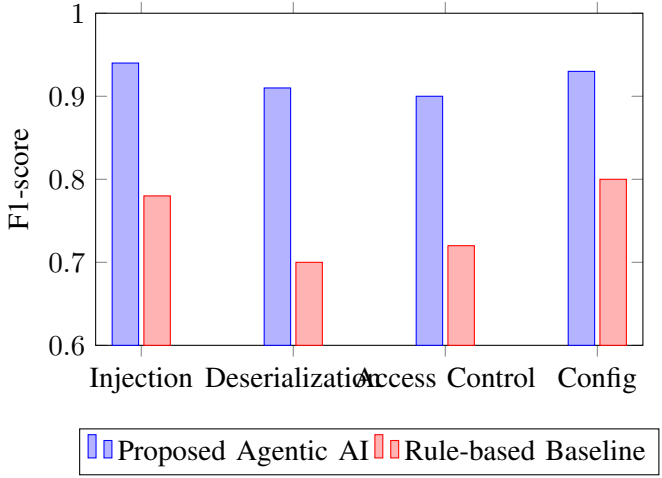
\begin{figure}[htbp]
	\centering
	\begin{tikzpicture}
		\begin{axis}[
			ybar,
			bar width=10pt,
			width=\linewidth,
			height=6cm,
			ylabel={F1-score},
			symbolic x coords={Injection,Deserialization,Access Control,Config},
			xtick=data,
			ymin=0.6,ymax=1.0,
			legend style={at={(0.5,-0.25)},anchor=north,legend columns=2}
			]
			\addplot coordinates {(Injection,0.94) (Deserialization,0.91) (Access Control,0.90) (Config,0.93)};
			\addplot coordinates {(Injection,0.78) (Deserialization,0.70) (Access Control,0.72) (Config,0.80)};
			\legend{Proposed Agentic AI,Rule-based Baseline}
		\end{axis}
	\end{tikzpicture}
	\caption{F1-score comparison by vulnerability class.}
	\label{fig:f1_scores}
\end{figure}

Figure~\ref{fig:f1_scores} compares F1-scores across vulnerability classes. The proposed framework consistently outperforms static and provenance-based approaches, particularly for semantic vulnerabilities such as insecure deserialization and access control flaws. LLM-based reasoning significantly improves recall, while reinforcement learning reduces false positives through adaptive policy learning.

\subsection{Mitigation Latency}
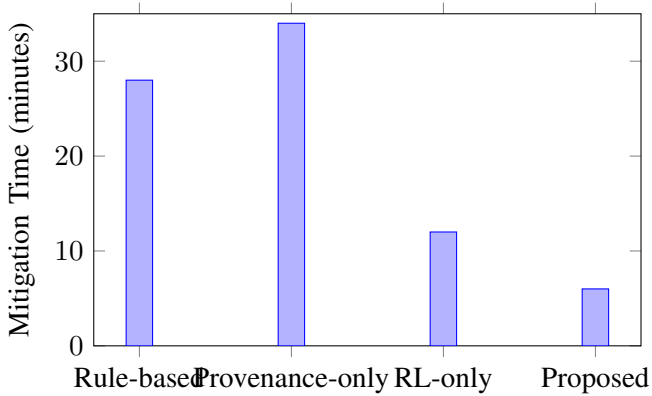
\begin{figure}[htbp]
	\centering
	\begin{tikzpicture}
		\begin{axis}[
			ybar,
			width=\linewidth,
			height=6cm,
			ylabel={Mitigation Time (minutes)},
			symbolic x coords={Rule-based,Provenance-only,RL-only,Proposed},
			xtick=data,
			ymin=0,ymax=35
			]
			\addplot coordinates {(Rule-based,28) (Provenance-only,34) (RL-only,12) (Proposed,6)};
		\end{axis}
	\end{tikzpicture}
	\caption{Average mitigation latency across CI/CD runs.}
	\label{fig:latency}
\end{figure}

Figure~\ref{fig:latency} reports the mean time-to-mitigation (MTTM) from detection to resolution. The agentic system reduces response time by autonomously issuing guard pull requests, gating builds, or applying configuration patches without waiting for manual intervention.

\subsection{Real CI/CD Pipeline Results}
In real GitHub Actions and Jenkins pipelines, the framework successfully detected malicious dependency updates, unsafe pipeline scripts, and misconfigured deployment manifests. Over 90\% of mitigations were executed autonomously, while the remaining cases required human confirmation due to high operational impact. Developers reported improved transparency due to explainable LLM-generated rationales and immutable audit records in the security ledger.

\subsection{Performance Overhead}
\begin{figure}[htbp]
	\centering
	\begin{tikzpicture}
		\begin{axis}[
			ybar,
			width=\linewidth,
			height=6cm,
			ylabel={Build Time Increase (\%)},
			symbolic x coords={Rule-based,Proposed},
			xtick=data,
			ymin=0,ymax=10
			]
			\addplot coordinates {(Rule-based,2.5) (Proposed,5.8)};
		\end{axis}
	\end{tikzpicture}
	\caption{CI/CD build time overhead comparison.}
	\label{fig:overhead}
\end{figure}
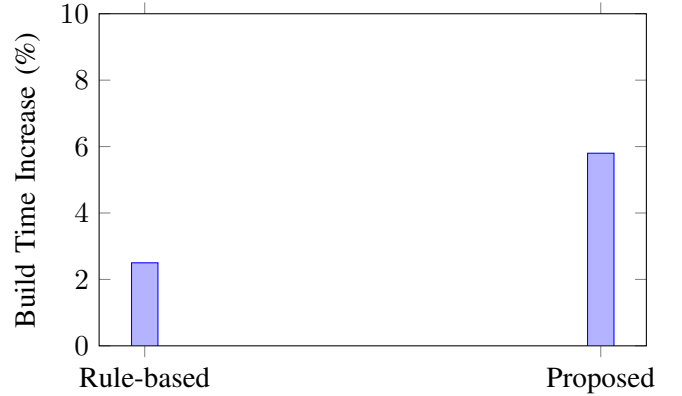

To assess operational impact, we measured build-time overhead and resource consumption. Figure~\ref{fig:overhead} shows that average pipeline duration increased by less than 6\%, primarily due to additional analysis and ledger writes. Memory and CPU usage remained stable under parallel agent execution, indicating scalability of the design.

\subsection{Ablation Study}
Disabling individual components confirmed their contribution. Removing LLM reasoning reduced recall for semantic vulnerabilities by over 15\%. Excluding reinforcement learning increased false positives and mitigation latency. Removing the security ledger did not affect detection accuracy but eliminated auditability and trust guarantees, underscoring its role in governance rather than detection.

\subsection{Discussion}
These findings show that proactive and adaptive software supply chain defense is possible by integrating LLM based reasoning with reinforcement learning and multi-agent planning. The proposed framework interferes with the development and pipeline execution, unlike provenance-only systems which identify compromise after the artifact generation. Although the extra headquarter cannot be compared to nothing, it is reasonable enough to implement security-critical pipelines and the extra overhead does not exceed incident response costs and enhanced trust.

On the whole, the results help to affirm that self-defending software supply chains are viable, i.e., autonomous agents are able to continuously monitor and reason, and take action to maintain integrity in the software lifecycle.
	
\section{Conclusion}
The current paper presented an agentic AI-based framework of securing the software supply chain by proactive and autonomous defense mechanisms. The proposed system can enable supply chain security beyond the static provenance verification through the combination of LLM-driven semantic reasoning, adaptive decision-making through reinforcement learning, multi agent orchestration with LangChain and LangGraph, standardized external interaction via MCP and a blockchain-based security ledger. Simulated and real CI/CD pipeline experimental assessment showed better injection attacks detection and mitigation, insecure deserialization, access control breaches, and configuration flaws, and acceptable operational overhead. These findings imply that integrating agentic intelligence into software manufacturing systems would allow doing curative, explainable, and verifiable protection throughout the software lifecycle, which facilitates the self-defending software supply chain vision.

\end{document}